\documentclass[12pt,letter,useAMS,natbib]{emulateapj-rtx4}
\usepackage{graphicx,amssymb}
\usepackage{txfonts}
\usepackage{natbib}

\bibpunct[, ]{(}{)}{;}{a}{}{,}
\def \aj {AJ}
\def \mnras {MNRAS}
\def \pasp {PASP}
\def \apj {ApJ}
\def \apjs {ApJS}
\def \apjl {ApJL}
\def \aap {A\&A}
\def \nat {Nature}
\def \araa {ARAA}

\newcommand{\kms} {$\mathrm{ km \; s^{-1}}\,$}

\def\lesssim{\mathrel{\hbox{\rlap{\hbox{\lower4pt\hbox{$\sim$}}}\hbox{$<$}}}}
\def\gtrsim{\mathrel{\hbox{\rlap{\hbox{\lower4pt\hbox{$\sim$}}}\hbox{$>$}}}}

\long\def\symbolfootnote[#1]#2{\begingroup%
\def\thefootnote{\fnsymbol{footnote}}\footnote[#1]{#2}\endgroup} 
\begin{document}


\title{The Yellow Supergiant Progenitor of the Type II Supernova 2011dh in M51}
\author{J. R. Maund\altaffilmark{1,2,3}, M. Fraser\altaffilmark{4},  M. Ergon\altaffilmark{5}, A. Pastorello\altaffilmark{6}, S.J. Smartt\altaffilmark{4}, J. Sollerman\altaffilmark{5}, S. Benetti\altaffilmark{6}, M.-T. Botticella\altaffilmark{6}, F. Bufano\altaffilmark{7}, I.J. Danziger\altaffilmark{8}, R. Kotak\altaffilmark{4}, L. Magill\altaffilmark{4,9}, A.W. Stephens\altaffilmark{10}, S. Valenti\altaffilmark{6}}, 

\altaffiltext{1}{Dark Cosmology Centre, Niels Bohr Institute, University of Copenhagen, Juliane Maries Vej, DK-2100 Copenhagen \O, Denmark; justyn@dark-cosmology.dk}
\altaffiltext{2}{Department of Astronomy \& Astrophysics, University of California, Santa Cruz, 95064, U.S.A.}
\altaffiltext{3}{Sophie \& Tycho Brahe Fellow}
\altaffiltext{4}{Astrophysics Research Center, School of Mathematics and Physics, Queen’s University Belfast, Belfast, BT7 1NN, Northern Ireland}
\altaffiltext{5}{Oskar Klein Centre, Department of Astronomy, AlbaNova, Stockholm University, 106 91 Stockholm, Sweden}
\altaffiltext{6}{INAF-Osservatorio Astronomico di Padova, Vicolo dell'Osservatorio 5, 35122 Padova, Italy}
\altaffiltext{7}{INAF-Osservatorio Astronomico di Catania, Via S.Sofia 78, 95123 Catania, Italy}
\altaffiltext{8}{INAF-Osservatorio Astronomico di Trieste, Via G.B. Tiepolo 11, 34131 Trieste, Italy}
\altaffiltext{9}{Isaac Newton Group of Telescopes, Apartado de Correos 321, E-38700 Santa Cruz de la Palma, Tenerife, Spain}
\altaffiltext{10}{Gemini Observatory, 670 North Aohoku Place, Hilo, HI 96720, USA}

\begin{abstract}
  We present the detection of the progenitor of the Type II SN~2011dh
  in archival pre-explosion Hubble Space Telescope images.  Using
  post-explosion Adaptive Optics imaging with Gemini NIRI+ALTAIR, the
  position of the SN in the pre-explosion images was determined to
  within $23$mas.  The progenitor object was found to be consistent
  with a F8 supergiant star ($\mathrm{log} L/L_{\odot}=4.92\pm0.20$
  and $T_{eff}=6000\pm280K$).  Through comparison with stellar
  evolution tracks, this corresponds to a single star at the end of
  core C-burning with an initial mass of $M_{ZAMS}=13\pm3M_{\odot}$.
  The possibility of the progenitor source being a cluster is
  rejected, on the basis of: 1) the source is not spatially extended;
  2) the absence of excess $\mathrm{H\alpha}$ emission; and 3) the
  poor fit to synthetic cluster SEDs.  It is unclear if a binary
  companion is contributing to the observed SED, although given the
  excellent correspondence of the observed photometry to a single star
  SED we suggest the companion does not contribute significantly.
  Early photometric and spectroscopic observations show fast evolution
  similar to the transitional Type IIb SN~2008ax, and suggest that a
  large amount of the progenitor's hydrogen envelope was removed
  before explosion.

\end{abstract}
\keywords{supernovae:general -- supernovae:individual (2011dh)}

\section{Introduction}
\label{sec:intro}

The search for the progenitors of core-collapse supernovae (CCSNe) has
now become an integral part of the study and understanding of nearby
supernovae (SNe). The last decade has shown that the global archives
of high resolution images of nearby galaxies can provide definitive
detections of progenitor stars on images of galaxies gathered before
explosion (for a review, see \citealt{2009ARA&A..47...63S}). The
luminosity and inferred mass of progenitors give an insight into the
origin of the SN types, explosion mechanisms and the last stages of
stellar evolution.\\
The most common type by volume, the Type II Plateau (IIP) SNe
\citep{2011MNRAS.412.1473L} have been shown to originate from red
supergiants
\citep{2003PASP..115.1289V,smartt03gd,2005MNRAS.364L..33M,2006ApJ...641.1060L}.
The disappearance of these stars several years after explosion is
further reassurance that they were indeed the stellar progenitors
\citep{2009Sci...324..486M}.  However the low and high mass end of the
distribution of progenitor masses are providing some unexpected
results.  The lack of high mass stellar progenitors is becoming
statistically significant \citep{2008arXiv0809.0403S}, and the
question of what SNe these high mass stars ($M_{\rm ZAMS} >
20M_{\odot}$) produce is critical to understanding stellar evolution,
black hole and neutron star formation and galactic chemical evolution.
While a number of high-mass progenitors have been associated with Type
IIn SNe \citep{galyam05gl,2011ApJ...732...63S}, the frequency of
high-mass progenitor detection is not high enough to comfortably match
the expected numbers from initial mass function considerations
\citep{2008arXiv0809.0403S}.  At the low luminosity end, faint
progenitors have been frequently detected
\citep{2005MNRAS.364L..33M,2006ApJ...641.1060L,2008ApJ...688L..91M,2010arXiv1011.6558F}.
They are often associated with the sub-luminous, low kinetic energy
and $^{56}$Ni-poor SNe which resemble SNe~2005cs and 1997D
\citep{andrea05cs}.  The obvious interpretation is that these are low
mass stars ($7-9M_{\odot}$) producing explosions of low energy
$\sim10^{50}$ ergs.\\
Several observers independently reported discoveries of SN~2011dh in
M51 \citep{CBET2736}, with the earliest recorded epoch being May
31.954, at a position $\mathrm{\alpha_{J2000}=13^{h} 30^{m} 05^{s}.12
  \, , \, \delta_{J2000}= +47\degr 10\arcmin 11\arcsec}$.
\citet{CBET2736a} and \citet{CBET2736b} classified it as a young Type
II SN.  The Whirlpool galaxy has now hosted three modern SNe (1994I,
2005cs and 2011dh), and the Hubble Heritage images
\citep{2005AAS...206.1307M} provide deep multi-color images for a
progenitor star search, as already done for SN~2005cs
\citep{2005MNRAS.364L..33M,2006ApJ...641.1060L}.  This letter reports
the discovery of the progenitor of SN~2011dh using Gemini North
NIRI+ALTAIR Adaptive Optics (AO) images of the SN and archival
pre-discovery optical  images.  We adopt a
distance of 7.1$\pm1.2$\,Mpc \citep{2006MNRAS.372.1735T}, and a
recessional velocity of $600$\,\kms from
NED\footnote{http://ned.ipac.caltech.edu/}.  The foreground reddening
towards M51 as quoted by NED, after \citet{schleg98}, is $E(B - V ) =
0.035$.  Using the metallicity gradient derived by
\citet{2004ApJ...615..228B} for {\sc H ii} regions in M51, we adopt a
metallicity at the SN position of $\sim 0.9 \pm 0.1 Z_{\odot}$.

\section{ The Progenitor Star}
\label{sec:obs}
The site of SN~2011dh was observed, prior to explosion, with the
Hubble Space Telescope (HST) Wide-Field Planetary Camera 2 (WFPC2) and
the Advanced Camera for Surveys (ACS) Wide Field Channel (WFC)
instruments, on 2005 Nov 13 and 2005 Jan 20-21.  ACS observations were
conducted with the $F435W$, $F555W$, $F658N$ and $F814W$ filters (with
observations with each filter being composed of four dithered
exposures) of total duration 2720, 1360, 2720 and 1360s, respectively.
A WFPC2 observation from 2005 Nov 13 used the $F336W$ filter, composed
of two equal exposures (1300s).  Additional WFPC2 observations of the
site, acquired on 2001 Jun 9, used the $F450W$, $F555W$ and $F814W$
filters with total exposure times of 2000s per filter. For both sets
of WFPC2 observations, the site of the SN fell on the WF2 chip with
pixel scale 0.1\arcsec. 

The HST data were retrieved from the HST
archive\footnote{http://archive.stsci.edu/hst/search.php}.  The ACS
images were combined using the {\tt multidrizzle} task running under
{\sc PyRAF} to correct for the geometric distortion of the ACS WFC
chip and improve sampling of the point spread function (PSF),
resulting in a final pixel scale of $0.025\arcsec$.  Photometry of
these images was conducted using {\sc DAOphot}, with PSFs derived from
the data themselves.  Aperture corrections were derived
to correct the photometry to an aperture of radius $0.5$\arcsec, with
the additional correction to an infinite aperture using the tabulated
values of \citet{acscoltran}.  No correction was applied for charge
transfer inefficiency in the drizzled images.  In parallel, the four
distorted images for each filter were processed separately using the
{\sc DOLphot}
package\footnote{http://americano.dolphinsim.com/dolphot/}, which
utilises pre-computed PSFs and does include corrections for charge
transfer inefficiency.  The photometry from these two techniques was
found to agree to within the photometric errors.  The WFPC2 images
were processed and analysed using the {\sc HSTphot} package
\citep{dolphhstphot}.  

We observed SN~2011dh in the $K$-filter with NIRI and the ALTAIR AO
system on the Gemini North Telescope on 2011 Jun 6. The f/32 camera
was used, which gives 0.022\arcsec pixels over a
22\arcsec$\times$22\arcsec field of view. As the SN was bright, it was
used as a natural guide star for ALTAIR. Separate off-source frames
were taken to remove the sky background. The data were reduced using
the {\sc IRAF gemini} {\sc niri} package.  The final image has a
coadded exposure time of 3000s, and a full-width at half-maximum of
0.2 \arcsec (after binning by 2 pixels in both the x and y
directions).

\begin{figure*}
\includegraphics[width=18cm]{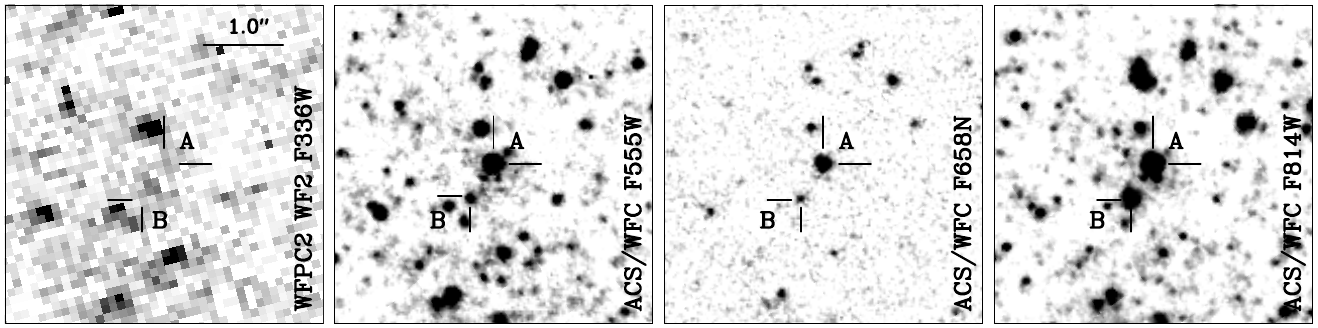}
\caption{Pre-explosion observations of the site of SN~2011dh.  Each
  panel has dimensions $4$\arcsec $\times$ $4$\arcsec, and is oriented
  such that North is up and East is left.  The progenitor candidate is
  denoted Source A and a nearby red star is denoted as Source B.
  From left to right, the panels are: Pre-explosion WFPC2 WF2 $F336W$
  image, pre-explosion ACS WFC F555W image, pre-explosion ACS WFC
  $F658N$ image, and pre-explosion ACS WFC $F814W$ image.}
\label{fig:obs:panel}
\end{figure*}

Geometric transformations between the post-explosion NIRI image and
the pre-explosion HST ACS and WFPC2 images were calculated with the
{\sc IRAF} task {\sc geomap}.  18 stars were identified in both the
pre-explosion ACS $F814W$ image and post-explosion NIRI image,
resulting in a transformation with rms error of 23mas.  We measured
the position of the SN in the NIRI image using the three different
centering algorithms in {\sc phot}. The standard deviation of the
three measurements is 2 mas. Transforming the coordinates of the SN as
measured in the NIRI image to the ACS $F814W$ image, we find the SN
position to be coincident with a bright, compact source as shown on
Figure \ref{fig:obs:panel} which we denote Source A.  This progenitor
candidate is located 3mas from the transformed position of the SN
(within the uncertainties of the transformation).

In the WFPC2 pre-explosion $F336W$ image, the progenitor was detected
at $4.4\sigma$ with $m_{F336W}=23.39\pm0.25$ (in flight system Vega
magnitudes).  In the drizzled ACS WFC images, the progenitor was
measured to have $m_{F435W}=22.36\pm0.02$, $m_{F555W}=21.83\pm0.04$,
$m_{F658N}=21.28\pm0.04$ and $m_{F814W}=21.20\pm0.03$. The 2001 WFPC2
$F450W$, $F555W$ and $F814W$ images give similar magnitudes, approximately $0.1-0.2$ magnitudes brighter than the ACS images (to be
expected due to the blending of the progenitor with a source to the
south).  There is no evidence for any variability in the progenitor's
brightness before explosion.  At the adopted distance to M51, after
correction for foreground reddening, this implies
$M_{F555W}=-7.54\pm0.37$.  

To determine if Source A, observed at the SN position in the
pre-explosion frames, was a cluster or a stellar object we employed
the {\sc ishape} package \citep{1999A&AS..139..393L} to probe its
spatial extent.  Due to the low S/N nature of the detection of the
source in the pre-explosion WFPC2 F336W image, as well as the
subsampled nature of the PSF in this image, we only analysed the
pre-explosion ACS/WFC images.  After subtracting nearby, well-resolved
stars from within a radius of 0.75\arcsec (or 1.5 times the radius of
the calculated PSF), {\sc ishape} was run on the images with both
delta and Moffat (order 1.5) functions convolved with the PSF.  {\sc
  ishape} was permitted to recalculate the coordinates of the source
and the source was permitted to be elliptical.  In all cases, the
shape of Source A was found to be consistent with an unresolved, point
source.
 
\begin{figure}
\includegraphics[width=8.5cm]{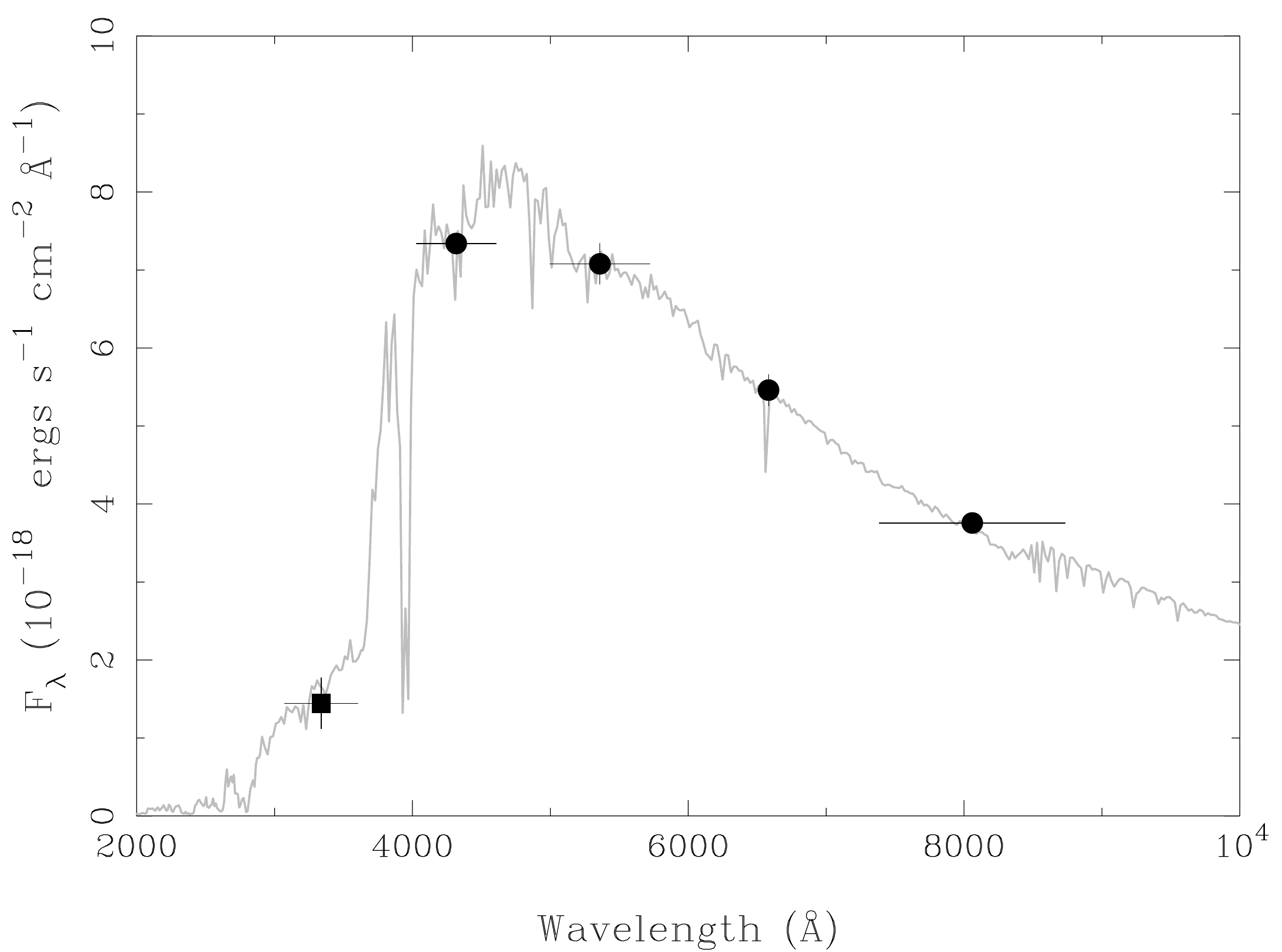}
\caption{The observed SED of the progenitor of SN 2011dh, as measured
  from pre-explosion HST WFPC2 ($\blacksquare$) and ACS/WFC
  ($\bullet$) images.  An ATLAS synthetic spectrum for a star with
  $T_{eff}=6000K$ and $\mathrm{log} (g) = 1.0$ is shown in grey.}
\label{fig:obs:sed}
\end{figure}

The observed photometry was compared with synthetic photometry for
model spectral energy distributions (SEDs) for single stars.  We used
{\sc chorizos} \citep{2004PASP..116..859M}, an SED fitting package,
with ATLAS synthetic spectra
\citep{1993KurCD..13.....K}\footnote{CHORIZOS and the SED library were
  obtained from http://jmaiz.iaa.es/software/chorizos/chorizos.html}
assuming a solar metallicity and a \citet{ccm89} $R_{V}=3.1$
reddening law.  The observed SED is shown on Fig. \ref{fig:obs:sed}.
The parameters derived for Source A are $T_{eff}=6000\pm280K$,
$\mathrm{log}(g) < 4$ and $E(B-V)=-0.01\pm0.1$ with $\chi^{2}=2.05$.
The {\sc chorizos} analysis was tested using a reddening-free index
$Q=(F555W - F658N)-0.947 \times (F658N-F814W)$, which was selected due to its
monotonic nature from 4000 to 8000K and being single valued for $T >
10\,000K$.  The temperature and the associated uncertainties calculated
from the reddening-free index, using the same ATLAS spectra, were
identical to those calculated using {\sc chorizos}.  In addition,
similar stellar parameters were calculated using MARCS synthetic
spectra \citep{2008A&A...486..951G}.

The photometry was also compared with the synthetic SEDs of clusters
generated using Starburst99 \citep{1999ApJS..123....3L} yielding
$\mathrm{log (age/years)}=9.9\pm0.1$ with $\chi^{2}=11.7$. This age is
significantly higher than the age of clusters observed around the SN
location ($\mathrm{log (age/years)} < 7.5$; \citealt{2009A&A...494...81S}).
This age is also inconsistent with the lifetimes of massive stars
$M_{ZAMS} > 8M_{\odot}$ that are expected to end their lives as SNe.
A further deficiency in interpreting the observed SED as that of a
cluster is the absence of $\mathrm{H\alpha}$ excess, with the observed
brightness in $F658N$ being consistent with stellar continuum (see
Fig. \ref{fig:obs:sed}).

\begin{figure}
\includegraphics[width=8.5cm]{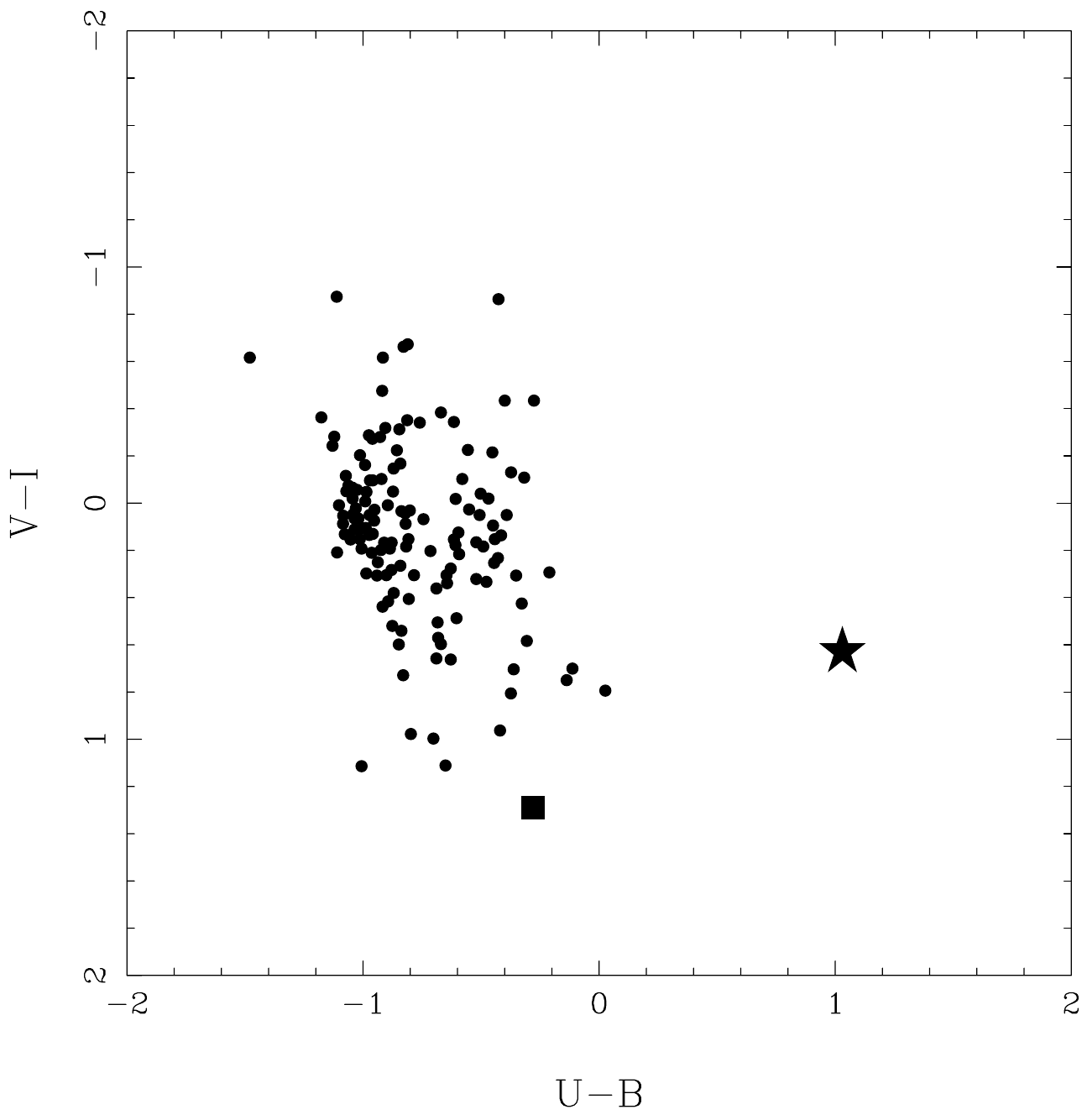}
\caption{The observed $U-B$ and $V-I$ colors of the progenitors of SNe
  2011dh ($\star$) and 1993J ($\blacksquare$), compared with the
  colors of WR stars ($\bullet$) in M33 \citep{2006AJ....131.2478M}.
  All colors have been corrected for foreground reddening.}
\label{fig:obs:wr}
\end{figure}

We compared the $U-B$ and $V-I$ colors of the progenitor candidate to
those of the observed population of Wolf-Rayet (WR) stars in M31
\citep{2006AJ....131.2478M}. All data have been corrected for
foreground Milky Way extinction. As can be seen in
Fig. \ref{fig:obs:wr}, the progenitor color is not consistent with a
WR star. Note that the progenitor colors are in the HST flight system,
which for the $F435W$, $F555W$ and $F814W$ ACS filters are consistent
(to within our photometric uncertainties) with the Johnson-Cousins
filters. In the case of the $U$-band, however, the difference between
the $F336W$ filter and the Johnson $U$ is non-negligible. Using
synthetic photometry of Potsdam WR model SEDs \citep{wolfspec}, we
find a $U-F336W$ color difference of $0.55\pm0.2$ mags. Applying this
to the progenitor, the $U-B$ color increases to 1.5 mags, which makes
the discrepancy between with the WR population even more marked.

\begin{figure}
\includegraphics[width=8.5cm]{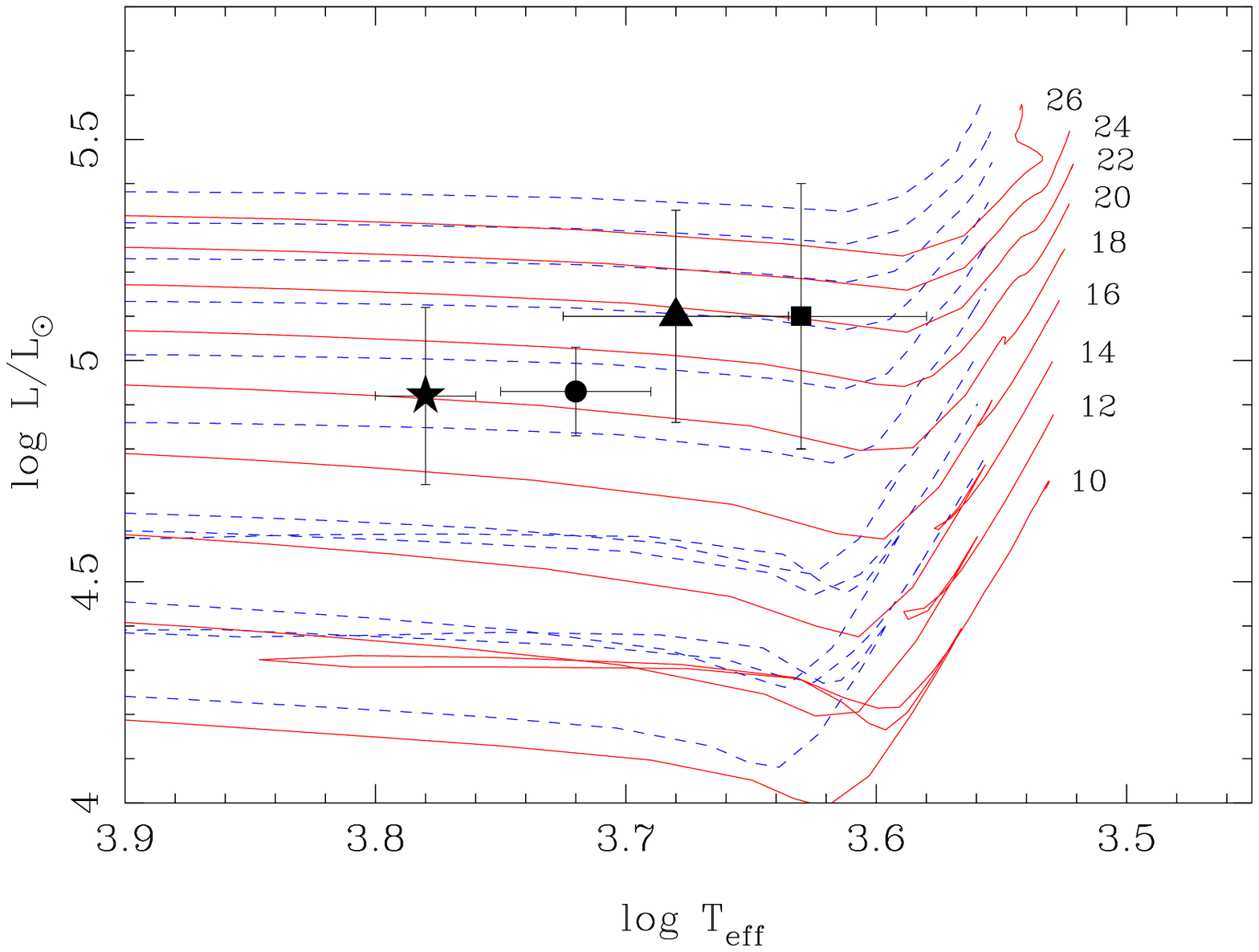}
\caption{Hertzsprung-Russell diagram showing the luminosities and
  temperatures of the progenitors of SNe 2011dh ($\star$), 1993J
  ($\blacksquare$; \citealt{maund93j,alder93j}), 2008cn ($\bullet$;
  \citealt{2009ApJ...706.1174E}), and 2009kr ($\blacktriangle$;
  \citealt{2010ApJ...714L.280F,2010ApJ...714L.254E}).  Overlaid are
  {\sc stars} stellar evolution tracks for solar (red solid) and LMC
  (blue dashed) metallicities.  At the end of each track the corresponding initial mass is indicated.}
\label{fig:obs:hrd}
\end{figure}

Given the point-like nature and colors of Source A, we conclude that
Source A is consistent with a F8 supergiant star with zero intrinsic reddening.
Utilising a bolometric correction of ($-0.013\pm0.074$) and a color
correction Johnson $V-F555W = -0.01\pm0.01$, derived from ATLAS
spectra, we infer a luminosity for Source A of $\mathrm{log} \left( L
  / L_{\odot} \right)=4.92\pm0.20$.  The location of the progenitor
object on the Hertzsprung-Russell (HR) diagram  is shown on
Fig. \ref{fig:obs:hrd} and compared with {\sc stars} stellar evolution
tracks \citep{eld04}.  In deriving the mass estimate, we use the final
luminosities for progenitor stars at the end of core C-burning \citep[see][and
their Fig. 1]{2008arXiv0809.0403S} which yields an initial mass of
$13\pm3M_{\odot}$.  The object we have called Source A is very likely
to be the same source identified by \citet{ATEL3401}, however they
derive an initial mass of $18-24M_{\odot}$.

\section{The early characteristics of SN 2011dh}
\label{sec:early}
\begin{figure*}
\includegraphics[width=18cm]{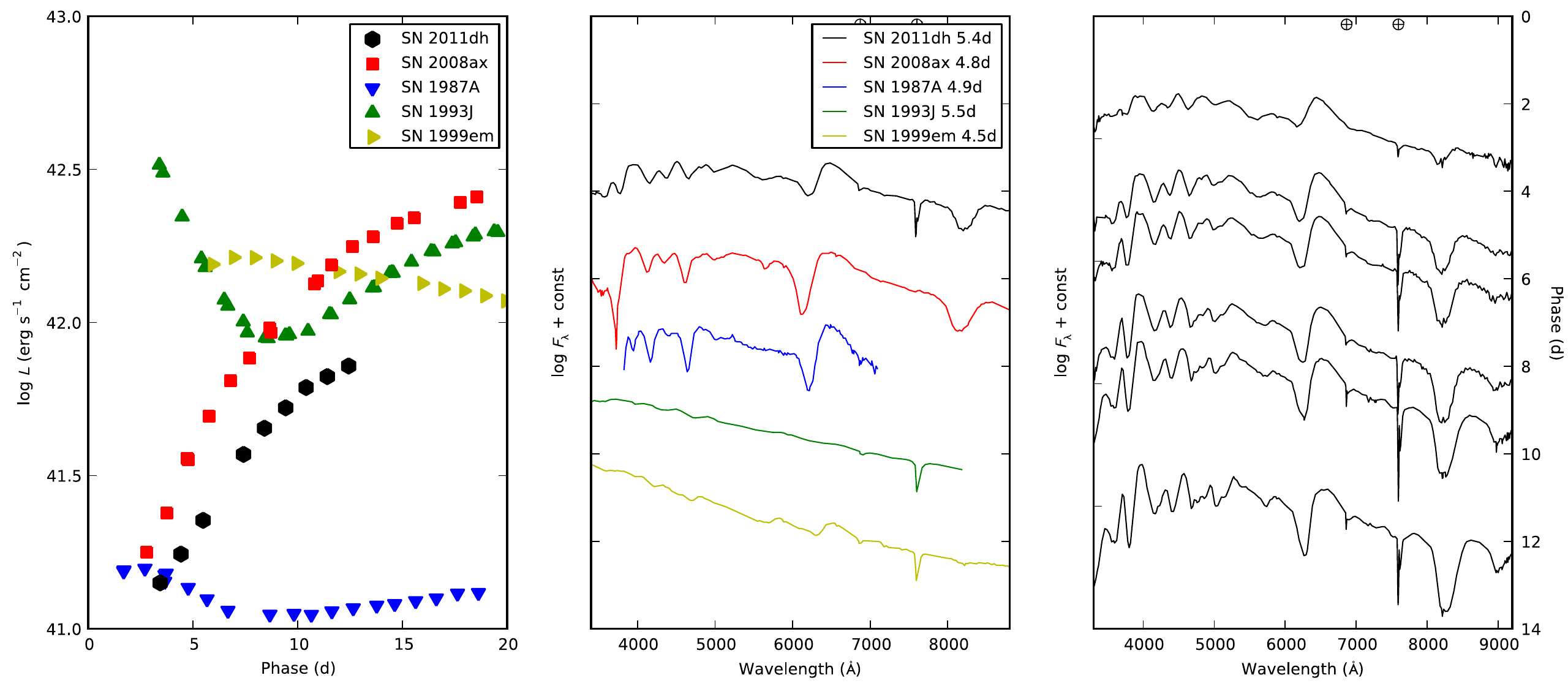}
\caption{{\it Left panel)} Pseudo-bolometric UBVRI lightcurves for SN
  2011dh and comparison SNe calculated as described in
  \citet{2010arXiv1011.6558F}. {\it Middle panel)} Optical spectra for
  SN 2011dh and comparison SNe at $\sim$5d. {\it Right panel)}
  Optical spectroscopic evolution for SN 2011dh. To visualise the
  temporal evolution the spectra have been aligned to the time axis at
  the right border of the panel. Telluric lines are indicated with a
  $\oplus$ symbol. All spectra have been corrected for redshift as
  given by NED. All spectra and photometry have been corrected for
  extinction using the extinction law of \citet{ccm89} and
  $R_{V}$=3.1. The phase is given in days relative to an assumed
  explosion date of 2011 May 31.5, estimated from reported detections
  and non-detections \citep{CBET2736}.}
\label{fig:sn}
\end{figure*}
Soon after the announcement of the SN discovery, a wide European
collaboration started a monitoring campaign using several telescopes
available to the collaboration. The results of the complete follow-up
campaign will be published in a forthcoming paper (Ergon et al. in
prep.). Here we present data obtained during the first $\sim$10
days after the explosion.

All images were bias, flat-field and overscan corrected.  The SN
photometry was measured using PSF-fitting and calibrated with
reference to several stars of the stellar sequence presented in
\citet{andrea05cs}. Spectroscopic data of SN 2011dh have been
processed using standard techniques, using the QUBA pipeline
\citep{2009jf}, and spectral fluxes were checked against the coeval
photometry.

Our data are presented in Figure \ref{fig:sn}, where we compare SN
2011dh to the Type IIb SNe 2008ax
\citep{2008MNRAS.389..955P,2011MNRAS.413.2140T} and 1993J
\citep{1994MNRAS.266L..27L}, the peculiar Type IIP SN 1987A
\citep{1987MNRAS.227P..39M,1988AJ.....95.1087P} and the normal Type
IIP SN 1999em \citep{2002PASP..114...35L, 2000ApJ...545..444B}. The
distance and extinction of SN 1987A are from
\citet{1990AJ.....99..650S}; the explosion epoch and extinction of SN
1999em is from \citet{2003MNRAS.338..939E}, while its distance is
taken from \citet{2003ApJ...594..247L}.

Contrary to the spectroscopic behavior of the Type IIb SN~1993J, the
early-time spectra of SN~2011dh show prominent H lines and relatively
weak He I features.  The light curve of SN 2011dh, however, has a rise
time markedly similar to that of the type IIb SN 2008ax \citep[see
also][]{ATEL3413}.  This is evidence against the presence of a
massive, extended H envelope. From the available photometry we
conclude that SN~2011dh is not a classical Type IIP SN, but further
monitoring is necessary to definitely associate SN 2011dh with one of
the other Type II subtypes (IIb, IIL, 1987A-like).

Our inference of negligible host galaxy reddening from our study of
the SED of the progenitor is supported by the non-detection of
\ion{Na}{1} D, at the rest-wavelength of M51, in high resolution
spectra acquired as part of our monitoring campaign (Ergon et al., in
prep.).

\section{Discussion \& Conclusions}
\label{sec:disc}
Despite the canonical prediction that Type II SNe arise from Red
Supergiants, there is mounting evidence that some stars explode as
Yellow Supergiants (YSGs).  A handful of Type II SNe have been
observed to arise from YSGs: SNe 1993J \citep{alder93j,maund93j},
2008cn (\citealt{2009ApJ...706.1174E}; Fraser et al., in prep.) and
2009kr \citep{2010ApJ...714L.280F,2010ApJ...714L.254E}.  The locations
of the progenitors on the HR diagram shows clearly that these stars are
not located on the predicted end points for single star stellar
evolution tracks.  In addition, despite arising from supposedly
similar YSG progenitors, these SNe display a wide range of properties.
While SN 1993J was a Type IIb SN (with most of its H envelope stripped
by mass transfer onto a binary companion), SN 2008cn was a bright Type
IIP SN.  \citet{2010ApJ...714L.254E} present evidence that SN~2009kr
is a Type IIL SN.  While \citet{ATEL3413} propose a Type IIb
classification for SN~2011dh, its early photometric and spectroscopic
properties (see Fig. \ref{fig:sn}) are not identical to SN~1993J.

The classification scheme for Type II SNe (IIP $\rightarrow$ IIL
$\rightarrow$ IIb) may be interpreted as being due to increasing mass
loss (i.e. stripping of the H envelope) from the progenitor.  The
effective temperatures derived for the YSG progenitors and the
classification of the resulting SNe shows this scheme is not
correlated with observed $T_{eff}$ of the progenitors
(Fig. \ref{fig:obs:hrd}).

It is clear from our observations that the SED of the progenitor of
SN~2011dh is consistent with a late-F supergiant.  While possible
combinations of two stars in a binary might also be employed to fit
the SED, any improvement in the quality of the fit is mitigated by two
key factors: 1) in introducing a binary companion, three additional
parameters (the temperature and gravity of the secondary and the ratio
of the brightness of the two stars at a reference wavelength) are
introduced into the fit; and 2) a companion star may not be
contributing any measurable flux to observed SED, yielding a single
star SED despite an underlying binary system. Unlike the case of
SN~1993J, we do not observe a UV excess associated with a possible
companion onto which mass from the progenitor of SN~2011dh may have
been transferred.  We propose that any binary companion is below the
detection limit of the pre-explosion observations.

The necessary binary parameters to produce such a YSG progenitor are also
unclear.  For example, the binary models of
\citet{2011A&A...528A.131C} only produce progenitor stars with
$T_{eff}<3860K$ and in that extreme case the companion is of similar
luminosity to the progenitor (with $T_{eff}=32\,200K$) that would
produce significant UV excess not observed in these pre-explosion images.

Despite the photometric and spectroscopic similarity to SN~2008ax, there
are significant differences between the proposed progenitor scenarios
for that SN \citep{2008MNRAS.391L...5C} and SN~2011dh.  While
\citeauthor{2008MNRAS.391L...5C} concluded the photometry of the
progenitor of SN~2008ax required either a single WR star or two
stellar components, our observations of the progenitor of SN~2011dh
are consistent with a single normal stellar component.

In isolation, very massive stars $>20M_{\odot}$ can lose significant
amounts of mass through stellar winds and eruptions (e.g. as Luminous
Blue Variables (LBV) and WR stars). We have demonstrated, however, that
the broad-band colors of the observed progenitor of SN~2011dh are
inconsistent with those of WR stars.  While some caution is required
in the application of this analysis to all YSG progenitors of Type II
SNe (e.g. the locus of the progenitor of SN~1993J, which was not a WR
star \citep{2009Sci...324..486M}, on Fig. \ref{fig:obs:wr}), that
caution is not justified for the progenitor of SN~2011dh.  As noted by
\citet{2010ApJ...714L.280F} for SN~2009kr, there is similarly no
evidence for previous LBV eruptions of the progenitor, as we do not
detect the signatures of eruptive mass loss such as: excess
$\mathrm{H\alpha}$ emission in the progenitor photometry or the
spectroscopic signature of the SN interacting with a dense
circumstellar medium (Fig. \ref{fig:sn}).

The nature of Source A as a single star or binary will be ultimately
confirmed at late-times, once the SN has disappeared and a surviving
binary companion, if present, is revealed again
\citep{maund93j,2009Sci...324..486M}.

\section*{Acknowledgements}
The research of JRM is funded through the Sophie \& Tycho Brahe
Fellowship.  The Dark Cosmology Centre is supported by the DNRF.
Based on observations obtained at the Gemini Observatory (program
GN-2011A-Q-22), which is operated by the Association of Universities
for Research in Astronomy, Inc., under a cooperative agreement with
the NSF on behalf of the Gemini partnership.  The data presented here
were obtained in part with ALFOSC, which is provided by the Instituto
de Astrofisica de Andalucia (IAA) under a joint agreement with the
University of Copenhagen and NOTSA.  Our thanks go to the staff of the
3.58m Telescopio Nazionale Galileo (La Palma, Spain), and to the
Asiago 1.82m Telescope (Asiago, Italy). SB and FB are partially
supported by the PRIN-INAF 2009 with the project "Supernovae Variety
and Nucleosynthesis Yields".  Based in part on observations obtained
with the Liverpool Telescope operated on the island of La Palma
(prog. XIL10B01). 
\bibliographystyle{apj}

\end{document}